\documentclass{webofc}
\usepackage[varg]{txfonts}   % Web of Conferences font
\usepackage{lineno}
\begin{document}
%\linenumbers
%
\title{The LHCspin project}
\subtitle{A polarised gas target at the Large Hadron Collider}

\author{\firstname{Marco} \lastname{Santimaria}\inst{1}\fnsep\thanks{\email{marco.santimaria@cern.ch}} 
        \and
        \firstname{Vittore} \lastname{Carassiti}\inst{2}
        \and
        \firstname{Giuseppe} \lastname{Ciullo}\inst{2,3}
        \and
        \firstname{Pasquale} \lastname{Di Nezza}\inst{1}
        \and
        \firstname{Paolo} \lastname{Lenisa}\inst{2,3}
        \and
        \firstname{Saverio} \lastname{Mariani}\inst{4,5}
        \and
        \firstname{Luciano Libero} \lastname{Pappalardo}\inst{2,3}
        \and
        \firstname{Erhard} \lastname{Steffens}\inst{6}
        % etc.
}

\institute{INFN - Laboratori Nazionali di Frascati, 
via Enrico Fermi 54, 00044 Frascati (RM), Italy
\and
INFN - Ferrara,
via Saragat 1, 44122 Ferrara (FE), Italy
\and
Department of Physics, University of Ferrara,
via Saragat 1, 44122 Ferrara (FE), Italy
\and
INFN - Firenze,
via Sansone 1, 50019 Sesto Fiorentino (FI), Italy
\and
Department of physics, University of Firenze,
via Sansone 1, 50019 Sesto Fiorentino (FI), Italy
\and
Physics department, FAU Erlangen-N\"{u}rnberg,
Staudtstr. 7 / B2, 91058 Erlangen, Germany
          }

\abstract{%
  The goal of LHCspin is to develop, in the next few years, innovative solutions and cutting-edge technologies to access spin physics in polarised fixed-target collisions at high energy, exploring the unique kinematic regime offered by LHC and exploiting new final states by means of the LHCb detector.
}
\maketitle
\section{Introduction}
\label{sec:intro}
The LHC delivers proton and lead beams with an energy of $7~\rm{TeV}$ and $2.76~\rm{TeV}$ per nucleon, respectively, with world's highest intensity.
Fixed-target proton-gas collisions occur at a centre-of-mass energy per nucleon of up to $115~\rm{GeV}$ in the case of a hydrogen target.
The large centre-of-mass boost offers an unprecedented opportunity to investigate partons carrying a large fraction of the target nucleon momentum, i.e. large Bjorken$-x$ values.

The LHCb detector~\cite{Alves:2008zz} is a general-purpose forward spectrometer specialised in detecting hadrons containing $c$ and $b$ quarks, and the only LHC detector able to collect data in both collider and fixed-target mode.
It is fully instrumented in the $2<\eta<5$ region with a vertex locator (VELO), a tracking system, two Cherenkov detectors, electromagnetic and hadronic calorimeters and a muon detector.

The fixed-target physics program of LHCb is active since the installation of the SMOG device~\cite{Aaij:2014ida}, enabling the injection of noble gases in the beam pipe section crossing the VELO detector at a pressure of $\mathcal{O}(10^{-7})~\rm{mbar}$.

\begin{figure}[ht]
\centering
\includegraphics[width=0.4\textwidth]{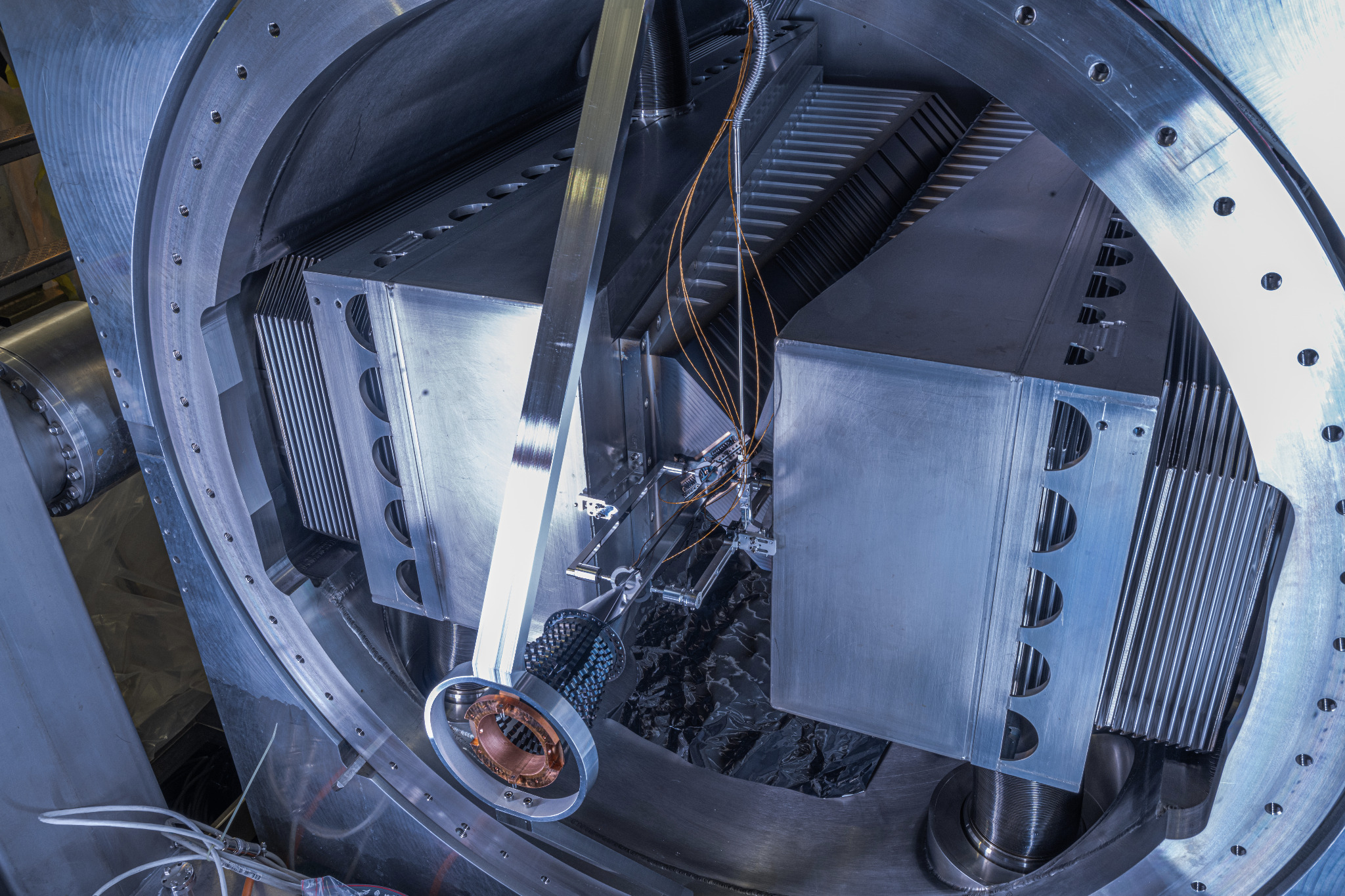}
\hfill
\includegraphics[width=0.58\textwidth]{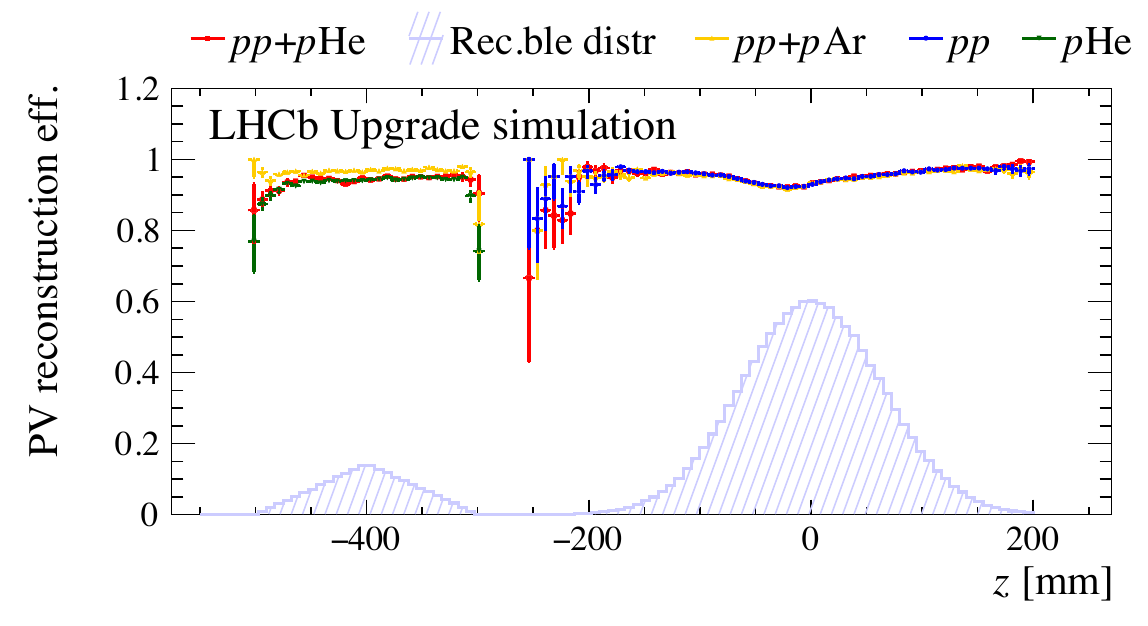}
\caption{Left: picture of the SMOG2 storage cell after installation. Right: simulated vertex reconstruction efficiency for simultaneous beam-gas and beam-beam collisions~\cite{LHCB-FIGURE-2019-007}.}
\label{fig:smog2}
\end{figure}

With the SMOG2 upgrade~\cite{LHCbCollaboration:2673690}, an openable gas storage cell, shown in Fig.~\ref{fig:smog2} (left), has been installed in 2020 in front of the VELO. 
The cell boosts the target areal density by a factor of $8$ to $35$ depending on the injected gas species, and creates a localised beam-gas collision region which is well detached from the proton-proton vertices. 
As shown in Fig.~\ref{fig:smog2} (right), full tracking efficiency is expected in the beam-gas region, despite its upstream position with respect to the VELO.
Due to these developments, fixed-target data will be collected in the upcoming Run 3 with a novel reconstruction software allowing for the simultaneous data-taking of beam-gas and beam-beam collisions. 
SMOG2 will offer a rich physics program for the Run 3 and, at the same time, will also allow to investigate the dynamics of the novel beam-target system, setting the basis for future developments.

The LHCspin project~\cite{Aidala:2019pit,10.21468/SciPostPhysProc.8.050} aims at extending the fixed-target physics program in Run 4 and Run 5 with the installation of a polarised gas target, bringing spin physics at LHC for the first time.
The experimental setup of LHCspin is discussed in Sec.~\ref{sec:det}, while a selection of physics opportunities is presented in Sec.~\ref{sec:phys}.

\section{Experimental setup}
\label{sec:det}
\begin{figure}[h]
\centering
\includegraphics[width=0.48\textwidth]{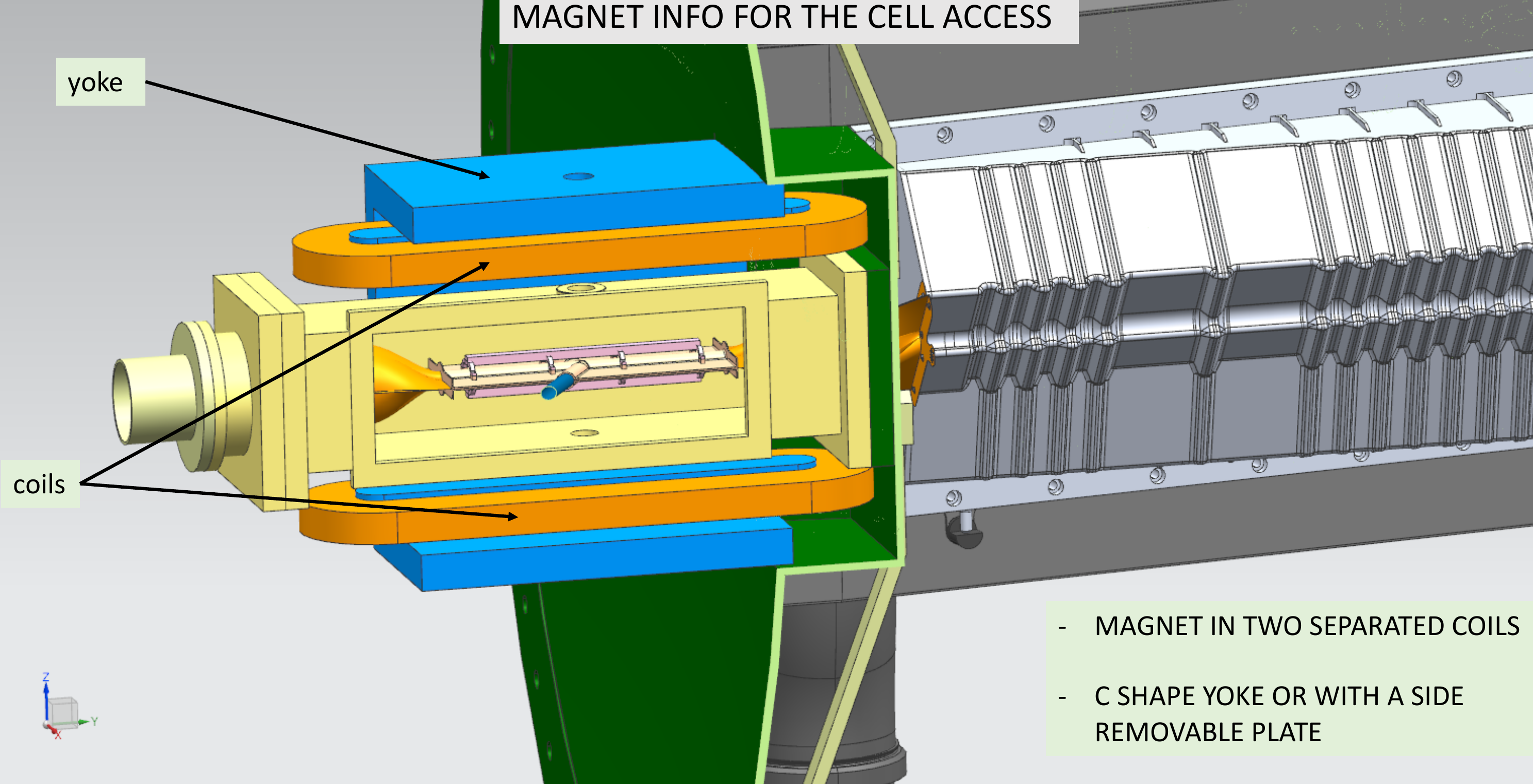}
\hfill
\includegraphics[width=0.5\textwidth]{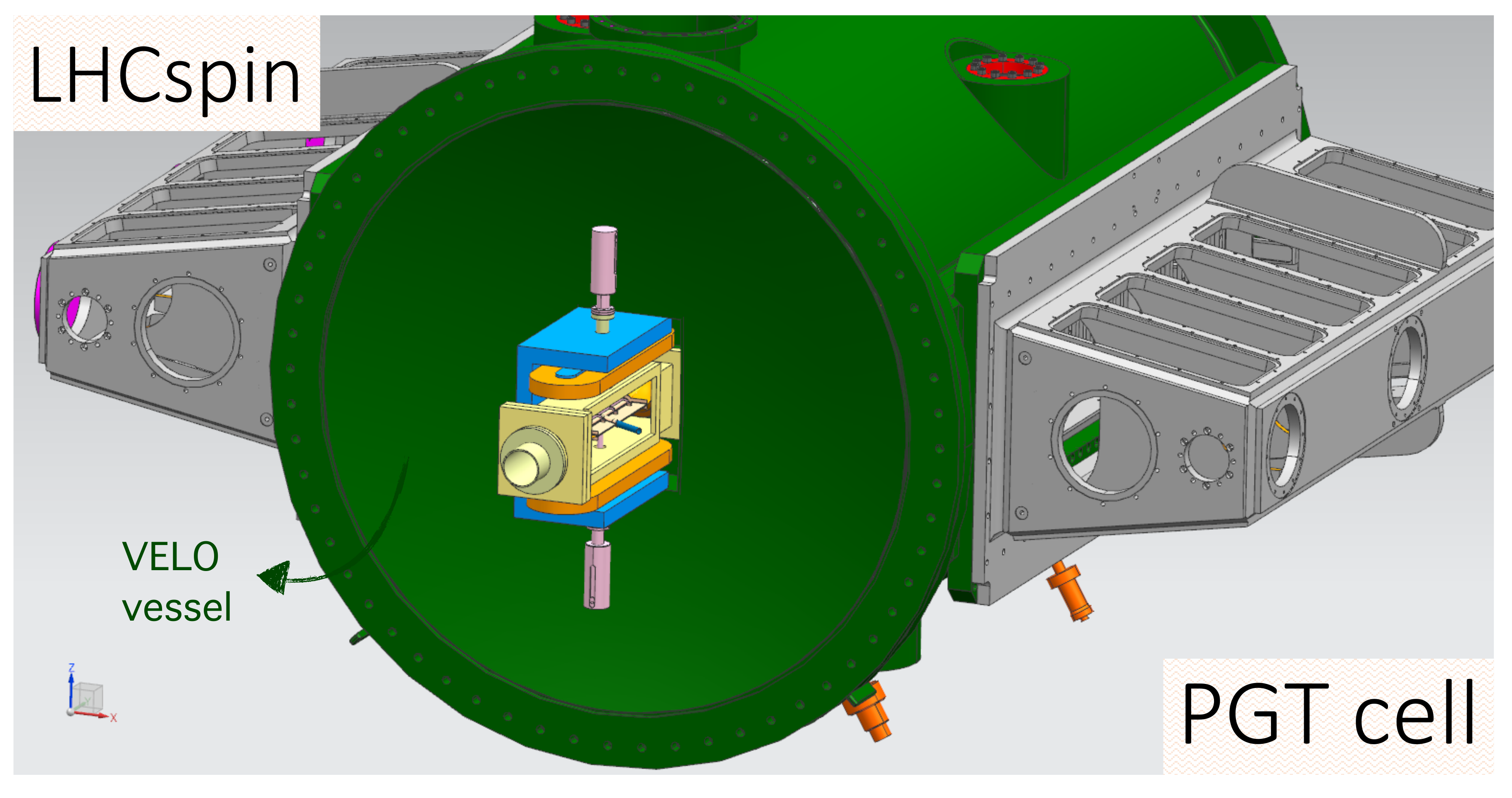}
\caption{Two views of the TC with the magnet coils (orange) and the iron return yoke (blue) enclosing the storage cell. The VELO vessel and detector box are shown in green and grey, respectively.}
\label{fig:rd}
\end{figure}

The LHCspin experimental setup is in R\&D phase and calls for the development of a new generation polarised target. The starting point is the setup of the HERMES experiment at DESY~\cite{Airapetian:2004yf} and comprises three main components: an Atomic Beam Source (ABS), a Target Chamber (TC) and a diagnostic system.
The ABS consists of a dissociator with a cooled nozzle, a Stern-Gerlach apparatus to focus the wanted hyperfine states, and adiabatic RF-transitions for setting and switching the target polarisation between states of opposite sign.
The ABS injects a beam of polarised hydrogen or deuterium into the TC, which is located in the LHC primary vacuum. The TC hosts a T-shaped openable storage cell, sharing the SMOG2 design, and a dipole holding magnet ($B=300$ mT), as shown in Fig.~\ref{fig:rd}.

The diagnostic system continuously analyses gas samples drawn from the TC and comprises a target gas analyser to detect the molecular fraction, and thus the degree of dissociation, and a Breit-Rabi polarimeter to measure the relative population of the injected hyperfine states. An instantaneous luminosity of $\mathcal{O}(10^{32})~\rm{cm}^{-2}\rm{s}^{-1}$ is foreseen for p-H collisions during Run 4.

\section{Physics case}
\label{sec:phys}
The physics case of LHCspin covers three main areas: exploration of the wide physics potential offered by unpolarised gas targets, investigation of the nucleon spin and heavy-ion collisions.

\paragraph{Unpolarised gas targets}
By means of the SMOG2 gas feed system, LHCspin will allow the injection of several species of unpolarised gases: H$_2$, D$_2$, He, N$_2$, O$_2$, Ne, Ar, Kr and Xe with negligible impact on the LHC beam lifetime. This gives an excellent opportunity to investigate parton distribution functions (PDFs) in both nucleons and nuclei in the large-$x$ and intermediate $Q^2$ regime, which is especially affected by lack of experimental data and impact several fields from basic QCD tests to astrophysics.
For example, the large acceptance and high reconstruction efficiency of LHCb on heavy flavour states enables the study of gluon PDFs, which are a fundamental input for theoretical predictions~\cite{Hadjidakis:2018ifr}.
Searches for an intrinsic charm component in the proton~\cite{Aaij:2018ogq} and antiproton production in p-He collisions~\cite{Aaij:2018svt} are two other high-profile example measurements that have already been pioneered at LHCb. 
With the large amount of data to be collected, nuclear PDFs can also be investigated in greater detail, helping to shed light on the intriguing anti-shadowing effect~\cite{Eskola:2016oht}, which is expected to be dominant in the $x$ range covered with LHCspin.

\begin{figure}[ht]
\centering
\includegraphics[width=0.48\textwidth]{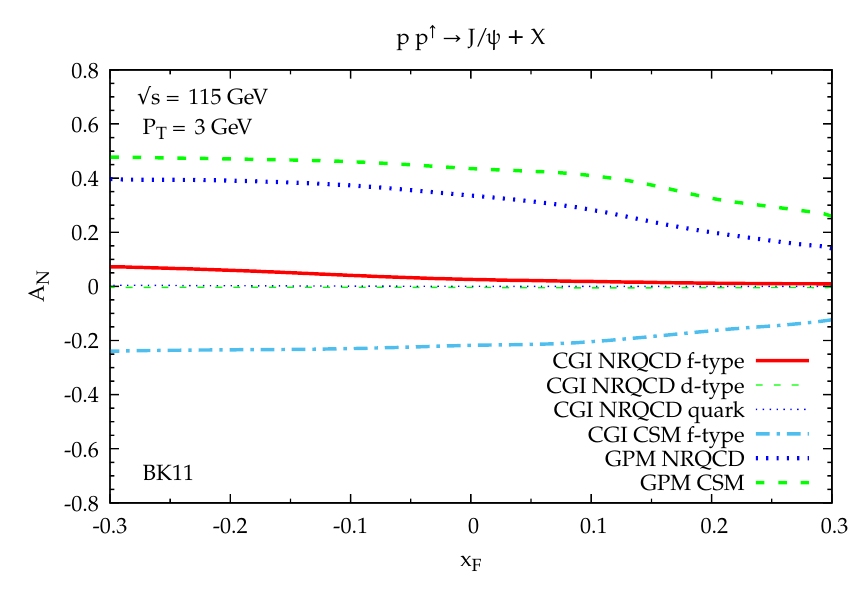}
\hfill
\includegraphics[width=0.5\textwidth]{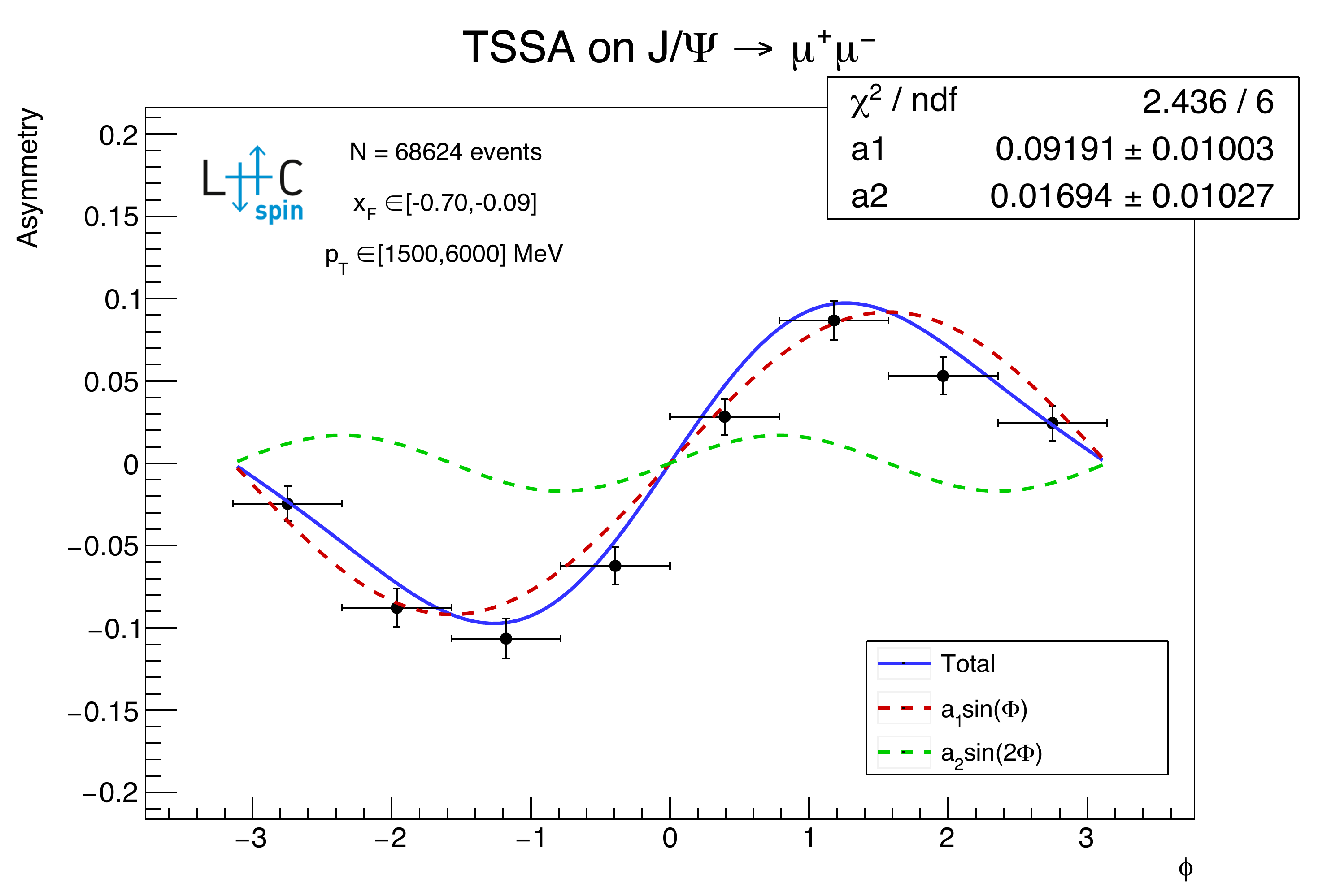}
\caption{Left: theoretical predictions for $A_N$ in inclusive $J/\psi$ production~\cite{PhysRevD.102.094011}. Right: simulated $J/\psi \to \mu^+\mu^-$ azimuthal asymmetries with a fit curve superimposed made of two Fourier terms.}
\label{fig:gsf}
\end{figure}

\paragraph{Spin physics}
Beside standard collinear PDFs, LHCspin will offer the opportunity to probe polarised quark and gluon distributions by means of proton collisions on polarised hydrogen and deuterium.
For example, measurements of transverse-momentum dependent PDFs (TMDs) provide a map of parton densities in 3-dimensional momentum space.
Light quark TMDs, especially in the high-$x$ regime, can be accessed by measuring transverse single spin asymmetries (TSSAs) in Drell-Yan processes, while gluon densities, such as the gluon Sivers function, can be probed via heavy-flavour production. Such asymmetry can be large (Fig.~\ref{fig:gsf}, left) and well in reach with just 1 month of LHCspin data, as shown in simulated $J/\psi \to \mu^+\mu^-$ events (Fig.~\ref{fig:gsf}, right).

It is also attractive to go beyond a 3-dimensional description by building observables which are sensitive to Wigner distributions~\cite{Bhattacharya:2017bvs} and to measure the elusive
transversity PDF, whose knowledge is currently limited to valence quarks at the leading order~\cite{Radici:2018iag}, as well as its integral, the tensor charge, which is of direct interest in constraining physics beyond the Standard Model~\cite{Courtoy:2015haa}. 

\paragraph{Heavy-ion collisions}
Thermal heavy-flavour production is negligible at the typical temperature of few hundreds MeV of the system created in ultra-relativistic heavy-ion collisions. Quarkonia states ($c\overline{c}$, $b\overline{b}$) are instead produced on shorter timescales, and their energy change while traversing the medium represents a powerful way to investigate Quark-Gluon Plasma (QGP) properties. 

LHCb capabilities allow to both cover the aforementioned charmonia and bottomonia studies and to extend them to bottom baryons as well as exotic probes.  QGP phase diagram exploration at LHCspin can be performed with a rapidity scan, complementing RHIC's beam-energy scan, while flow measurements will greatly benefit from the excellent particle identification performance of LHCb on charged and neutral light hadrons.
An interesting topic joining heavy-ion collisions and spin physics is the dynamics of small systems which can be probed via ellipticity measurements in lead-ion collisions on polarised deuterons~\cite{Broniowski:2019kjo}.

%
% BibTeX or Biber users please use (the style is already called in the class, ensure that the "woc.bst" style is in your local directory)
\bibliography{main.bib}
%\begin{thebibliography}{}
%Journal Author, Journal \textbf{Volume}, page numbers (year)
%Book Author, \textit{Book title} (Publisher, place, year) page numbers
% etc
%\end{thebibliography}

\end{document}